\documentstyle[aas2pp4]{article}
\lefthead{Zhang et al.}
\righthead{Milli-second Oscillations in Aql X-1...}

\newcommand{\etal}{{\em et al.}}

\begin{document}

\title{Milli-second Oscillations in the Persistent and Bursting Flux 
       of Aql X-1 During an Outburst}

\author{W. Zhang, K. Jahoda, R. L. Kelley, T. E. Strohmayer\altaffilmark{1},
        J. H. Swank}
\affil{Laboratory for High Energy Astrophysics \\
       Goddard Space Flight Center \\
       Greenbelt, MD 20771}
\and
\author{S. N. Zhang\altaffilmark{1} }
\affil{Marshall Space Flight Center \\
       ES-84\\
       Huntsville, AL 35812}

\altaffiltext{1}{{\it also} Universities Space Research Association}

\begin{abstract}

The Rossi X-Ray Timing Explorer observed the soft X-Ray
transient Aql X-1 during its outburst in February and March 1997. We
report the discovery of quasi-periodic oscillations (QPOs) in its
persistent flux with frequencies in the range of 740 to 830 Hz, Q-value
of over 100, and a fractional RMS amplitude of $(6.8 \pm 0.6)$\%, and
nearly coherent oscillations (NCOs) during a Type-I burst with a
frequency of 549 Hz.  The frequency of the QPOs in the persistent flux
is correlated with the mass accretion rate on time scale of hours, but
not on time scale of days. This is most likely the manifestation in a
single source of the kHz QPO puzzle observed among many sources, i.e.,
on the one hand, individual sources show a correlation between the QPO
frequency and the inferred mass accretion rate, on the other hand, the
dozen or so sources with luminosities spanning two decades have
essentially the same QPO frequencies.  We propose that this
multi-valued QPO frequency and mass accretion rate correlation
indicates the existence of many similar regimes of the accretion disk.
These regimes, with a very similar energy spectrum and QPO frequency,
are distinguished from each other by the mass accretion rate or the
total X-ray flux.  The NCOs during the burst can be made almost
perfectly coherent by taking into account a large $\dot{\nu}$. This
strongly suggests that this frequency is related to the neutron star
spin frequency.  The large $\dot{\nu}$ is attributable to the expansion
or contraction of the neutron star photosphere during the burst.

\end{abstract}


\section{Introduction}

The launch of the Rossi X-Ray timing Explorer (RXTE) in December 1995
marked a new beginning for the study of compact objects and their accretion
disks.  Since then
quasi-periodic oscillations (QPOs) in the kilo-Hertz range (kHz QPOs
henceforth) have been discovered in the persistent flux of a number of
low mass X-Ray binaries (LMXBs), including both the Z and the atoll
sources, as well as sources that have not been classified.
For a comprehensive review of the literature on this subject, see van 
der Klis (1997). Although questions about the detailed
physical mechanism for the generation of these oscillations has
not been settled, it is clear that the kHz QPOs are related to 
the dynamic time scale near the neutron star surface, and the
NCOs likely to be the neutron star spin. 

One of the puzzling characteristics of the kHz QPOs is their similarity
across the large number of sources which cover two decades in
luminosity and probably also two decades in magnetic field. This
characteristic raises a serious question as to what are the roles that
the overall accretion rate and the neutron star magnetic field play in
determining the QPO frequencies and other characteristics.  It places a
severe constraint on any potential model for their generation
mechanism. Another puzzling characteristic of both the kHz QPOs and the
NCOs is that the inferred neutron star spin frequencies of these 
LMXBs are in a very narrow range in the vicinity of 300 Hz
(\cite{white_and_zhang}).  Both of these puzzles are hard to understand
in the context of the standard disk magneto-dynamic interactions
(\cite{ghosh_and_lamb}) without invoking some kind of correlation
between the magnetic field and the mass accretion rate across these
sources (\cite{white_and_zhang}), unless we accept the possibility that
either the inferred frequencies from the burst oscillations are not
directly related to the neutron spins or the disk magneto-dynamic interactions
picture we have at the present is significantly incomplete.

In this paper we report RXTE observations of Aql X-1 during its
February 1997 outburst.  We present analysis results of the kHz QPOs in
the persistent flux and the NCOs during a Type-I burst. We believe that
the data strongly indicate that there exist new regimes characterized
by different branches in the relationship between the kHz QPO frequency
and the total source luminosity.

\section{Observations}

The data analyzed for this paper results from 12 separate RXTE
observations or pointings at Aql X-1 covering the calendar period from
16 February through 10 March 1997.  Each observation is typically two
RXTE orbits in duration and results in about 8,000 s of good data
segmented into two to three pieces due to RXTE passages through the
South Atlantic Anomaly and Earth occult of Aql X-1. The typical
time gap between two consecutive observations is two days. The first
observation took place approximately a week after Aql X-1 passed the
peak luminosity and the last when the source had essentially reached
its quiescent state.  Two Type-I X-ray bursts occurred during these
observations, but only one of which has data with good time and energy
spectral resolution.  The other one occurred shortly after an Earth
occult before the instrument was completely ready for taking data.
In addition to the standard 1 and 2 data sets, we have 
event mode data with 64 energy channels and 128$\mu t$ ($\mu t$
= $2^{-20} s$) time resolution, and a burst catcher mode with 1
energy channel and 128$\mu t$ time resolution. During the Type-I burst
the event mode data exceeds the bandwidth between the PCA and the
spacecraft, therefore has gaps. We have used the burst catcher data to
perform the timing analysis reported below.

\section{Data Analysis and Results}

\subsection{Characteristics of kHz QPOs in the Persistent Flux}

\begin{figure}
 \vskip 2.5 in
 \caption{Upper panel: the light curve from the data reported for this paper.
	  Lower panel: Fractional RMS for data from each of the 36 RXTE
	  orbits. The four filled circles have error bars too small
	  to show. In deriving the upper limits we have assumed the QPO
	  frequency can spread over the range of 700 to 900 Hz.}
 \label{pca_lc_and_rms}
 \includegraphics{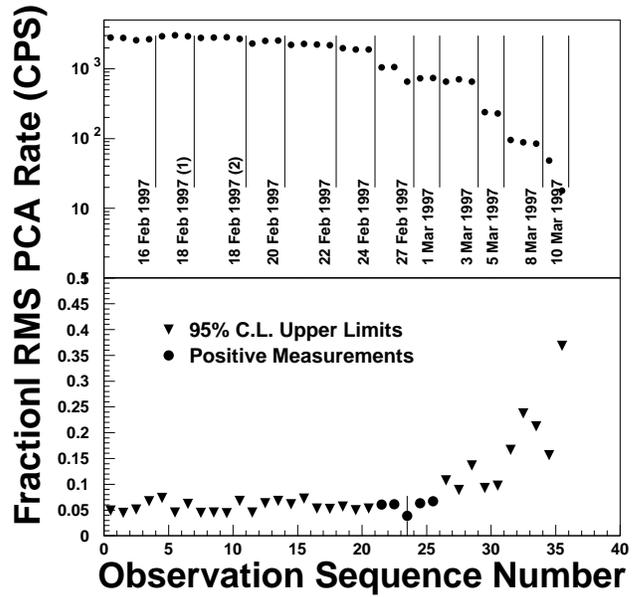}
\end{figure}

The upper panel of Figure~\ref{pca_lc_and_rms} shows the light curve
from the entire data set.  Each point represents data from one RXTE
orbit.  The 12 intervals bracketed by the vertical lines represent the
12 pointings.  We have fitted the energy spectrum with a model
consisting of a blackbody component and a power-law component with an
absorption column density fixed at $n_H=0.5\times 10^{22}$ cm$^{-2}$
(\cite{christian_and_swank}). Hereafter, flux values are calculated
using the fitted model and parameters.

\begin{figure}
 \vskip 2.5 in
 \caption{Power spectrum lined up with the detected QPO peak as 
	  the fiducial mark. The peak is at 760 Hz and
	  has a fractional RMS amplitude of
	  (6.8$\pm$0.6\%) and a FWHM of 6.3 Hz. There are no
	  peaks at any other frequencies.
	  The 95\% confidence level upper limit is 1\%.}
 \label{aqlx1_khz_lineup}
 \includegraphics{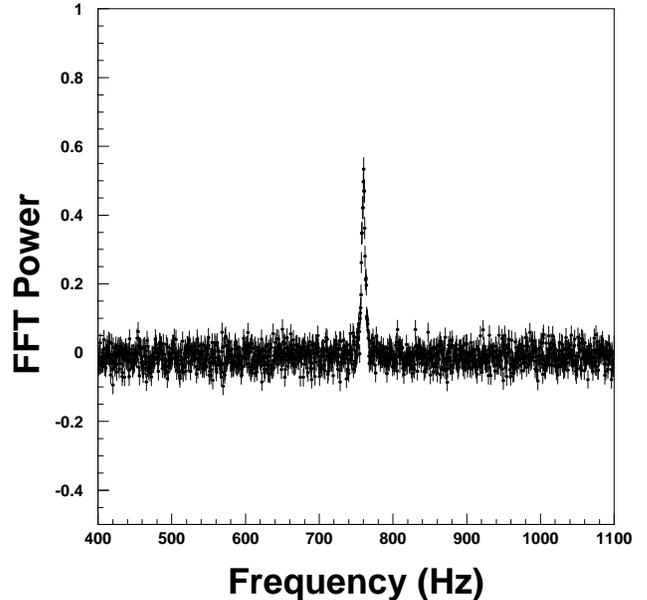}
\end{figure}

We have constructed average FFT power spectrum for data from every
orbit of the 12 pointings using event mode data.  Only 2 pointings have
shown clearly detectable kHz QPO peaks in their power spectra, the 7th
and the 8th which took place on 27 February and 1 March, respectively.
The lower panel of Figure~\ref{pca_lc_and_rms} shows the corresponding
upper limits for those days when kHz QPOs are not positively
detected.  The overall average fractional RMS amplitude in the
RXTE/PCA band is (6.8$\pm$0.6)\%. The width of the QPO peak varies with
time, but does not correlate with any other variable.  It has an
overall average of 6.23 Hz (FWHM) (see Figure~\ref{aqlx1_khz_lineup}).
Figure~\ref{freq_vs_flux} shows the correlation between the QPO
centroid frequency and the measured flux in the 2-10 keV band.  The
correlation between the frequency and count rate is very similar except
that there appears to be more scatter among the points. In particular
we have also examined the correlation between the QPO frequency and the
flux in the blackbody component and found that the correlation is much
worse. This is in contrast to the case of 4U 0614+091
(\cite{ford_etal_4u0614}) where the correlation between the two seem to
be rather good over a long time scale.  We note that  on each day
individually, the QPO frequency appears to be correlated with the flux,
but there is a shift between the two days. The 7 points toward the
lower left part of the graph which stand away from the rest of the
points are there because there was a 10\% drop in overall flux right
after the Type-I burst.  This kind of correlation on short time scale,
i.e., from minutes to hours, and the apparent lack of it on longer time
scale, i.e., days, between the QPO frequency and the overall flux or
count rate has been observed before in other sources, e.g., in 4U
1608-52 (\cite{4u1608_discovery,yu_etal_4u1608}) and 4U 0614+091
(\cite{ford_etal_4u0614}). It showed up very dramatically in these
observations of Aql X-1. The RMS amplitude varies with energy.
They are $(5 \pm 1)$\% and $(9 \pm 1)$\% in the 2-5 keV and 5-10 keV bands,
respectively.

Since only one QPO peak is observed on each of the two days, it is
possible that, even though they cover a similar range in frequency,
they could be the manifestations of the two QPO peaks that have been 
commonly observed in other sources, e.g., 4U 1728-34. To investigate 
this possibility, we have compared the dynamic FFT power spectrum of
4U 1728-34 obtained over a two day period in February 1996 and 
that of Aql X-1 on 27 February and 1 March 1997. We have found that 
in the case of 4U 1728-34, the QPO with the lower frequency is 
always much narrower in width than the QPO with the higher frequency.
They can be distinguished on power spectral width alone. On the other 
hand, the
QPO peaks of Aql X-1 during the two days look indistinguishable
in width. In particular, their widths are very similar to the
widths of the lower frequency QPO peak of 4U 1728-34. Therefore we
conclude that most likely the Aql X-1 QPOs we have observed during 
the two days are the same QPO peak.

Following Mendez \etal (1997) we have also searched
for a second QPO peak by fitting the power spectrum from every 64 s of
data and using the frequency from the fit to align all the power spectra.
As shown in Figure~\ref{aqlx1_khz_lineup} we have found no indication
of a second peak anywhere in the resulting power spectrum. In
particular we have examined the part of the power spectrum where one
would expect a second peak, namely 275 Hz above or below the existing
QPO peak and found no statistically significant enhancement at all. The
95\% confidence level upper limit is 1\%.

\subsection{Oscillations During the Burst}

\begin{figure}
 \vskip 2.5 in
 \label{burst_psd}    
 \caption{Power spectra obtained from the burst. Dashed histogram:
	  power spectrum obtained with the original times. Solid histogram:
	  power spectrum obtained after correcting the original times
	  with the $\dot{\nu}$.}
 \includegraphics{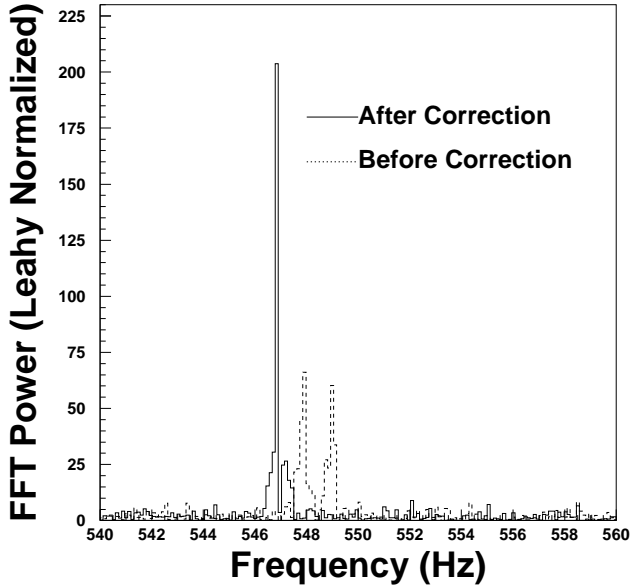}
\end{figure}

\begin{figure}
 \vskip 2.5 in
 \caption{Upper panel: Fractional RMS amplitude of the 549 Hz oscillations
	  in the burst vs. time. Lower panel: light curve of the
	  burst. The time zero in the plot corresponds to UTC 23:27:39
	  on 1 March 1997.}
 \label{burst_rms_vs_time}
 \includegraphics{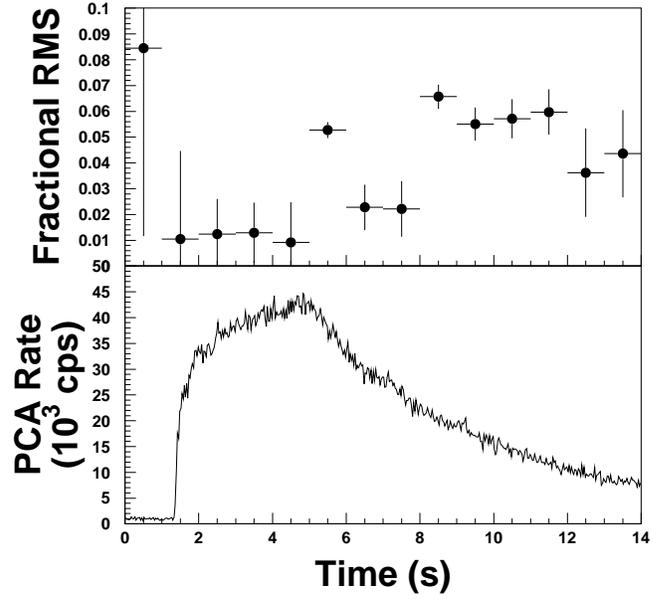}
\end{figure}

An FFT of the 14 s time series of the burst catcher data 
reveals that its intensity
oscillates at frequencies near 549 Hz, as shown by the dashed
histograms in Figure~\ref{burst_psd}. To investigate the frequency and
its amplitude variation with time, we have constructed power spectra
for each of the 14 seconds of data, as shown in
Figure~\ref{burst_rms_vs_time}. It can be seen that statistically
speaking, the oscillations do not start until after the peak of the
light curve which is shown in the lower panel.

To investigate whether the oscillation during the burst is coherent, we
have performed a search in $\nu - \dot{\nu}$ space for optimal values
of $\nu$ and $\dot{\nu}$ to minimize the width of the FFT peak.  The
detailed procedure is that, given a trial pair of ($\nu, \dot{\nu}$),
we first transform the original arrival time $t_0$ of each photon to
$t_1=t_0 + \dot\nu t_0^2 /\nu$ and then construct the FFT power
spectrum using the new time series of $t_1$.  Using the 8 s of data
right after the peak of the light curve, we have found that when
$\nu_0=547.8$ Hz and $\dot{\nu}$=0.14 Hz s$^{-1}$, we obtain the power
spectrum with the minimal width, as shown by the solid histogram in
Figure~\ref{burst_psd}.  The amount of power in the peaks of the dashed
and solid histograms is $418\pm 9$ and $425 \pm 9$, respectively.  In
other words, we have gathered all the powers from a much larger
frequency region into a much smaller region by taking into account a
$\dot{\nu}$.  The width of the resultant peak is such that, though it
is not consistent with perfect coherence, it is fairly close.
Therefore we conclude that this is strong evidence to support the
interpretation that the observed oscillations are due to the spin of
the neutron star photosphere. The large $\dot{\nu}$ indicates that the
photosphere is probably mostly decoupled from the rest of the star and
therefore conserves its own angular momentum in the course of
contraction (\cite{x1743_burst}).  The inclusion of a $\ddot{\nu}$ does
not decrease the width of the peak.

\section{Discussion}

\begin{figure}
 \vskip 2.5 in
 \caption{QPO frequency vs. flux in the 2-10keV band. The two
 clearly separated groups come from data collected on 27 February (right)
 and 1 March (left), respectively.}
 \label{freq_vs_flux}
 \includegraphics{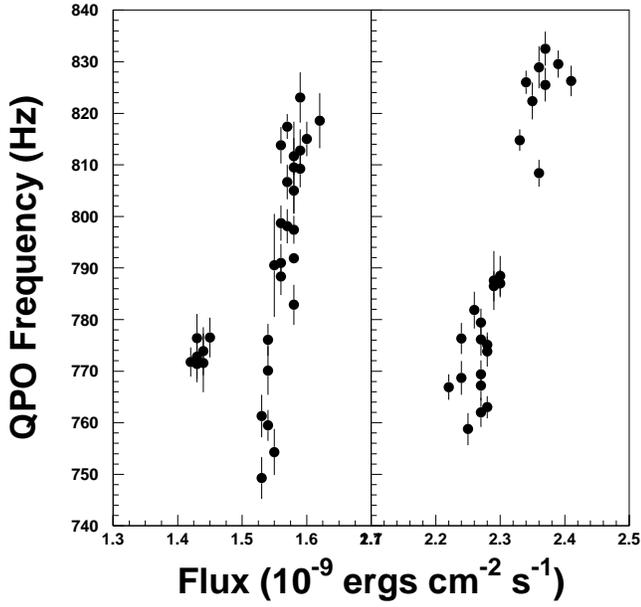}
			   
\end{figure}

Since the energy spectrum of the source changed very little, if any,
from 27 February to 1 March, it is quite reasonable to assume that the
measured flux and the true mass accretion rate are positively
correlated.  Figure~\ref{freq_vs_flux} clearly indicates that it is not
possible that, given a mass accretion rate, or for that matter, given
an X-ray flux level, the QPO frequency is uniquely determined. On the
contrary, Figure~\ref{freq_vs_flux} clearly indicates that at least two
flux levels can give the same QPO frequency. We believe this is a clear
manifestation in a single source of the puzzle we discussed in the
introduction, i.e., on the one hand, there appears to be a correlation
between the QPO frequency and the source luminosity for individual
sources; on the other hand, despite the vast difference in mass
accretion rates, the QPO frequencies are essentially the same among all
those sources.  It is therefore reasonable to conclude that,  in
LMXBs,  regimes with essentially identical QPO frequency and energy
spectrum can exist at very different fluxes or mass accretion
rates.

This conclusion is inevitable when one further examines
Figure~\ref{freq_vs_flux}. For example, on 27 February 1997, as the
flux decreased, the QPO frequency also decreased. Had the
observation been carried out continuously from 27 Feb through 1
March, we would have had to have observed one of the following three
possibilities: (1) at some point the kHz QPOs cease being observable
for some time and then becoming observable again at a higher
frequency; (2) at first the QPO frequency is positively correlated with
the flux until the frequency reaches a smaller value than 750 Hz, then
they become anti-correlated for some time so that as the flux
decreases, the QPO frequency increases until it reaches above 830 Hz
and the anti-correlation changes back to the positive correlation; (3)
as the flux decrease, there is a sudden and discontinuous jump in the
QPO frequency from below 750 Hz to above 830 Hz.  In all likelihood,
possibility (1) probably would have been observed.  The reason is that,
to our knowledge, the anti-correlation required by possibility (2) has
never been observed in any source at any time.  Neither has the sudden
and discontinuous change in QPO frequency of possibility (3).

The data we have presented have naturally led us to the
conclusion that the total X-ray flux, therefore the total mass accretion
rate, of a source can not uniquely determine the frequency of the kHz
QPOs. Moreover we infer from the data that there exist many
similar regimes for the inner region of the accretion disk. In each
of these regimes, the kHz QPO frequency is more or less tightly
correlated with the mass accretion rate or the total X-ray flux. The
QPO frequency ranges of these regimes are rather similar. They are all
in the vicinity of 1,000 Hz. The disk can transition from one regime to
another with change in the overall mass accretion rate. 

This transition of a LMXB from one state to another with change in mass
accretion rate is not new. For example, the Z source GX 5-1 has 
been believed to move along its Z track on the X-ray color-color
diagram as the mass accretion changes (\cite{hk89}).  This motion along
the Z track, however, involves significant changes in the energy
spectral shape. What is new with Aql X-1 is that, even though the
mass accretion rate and the energy spectral state can be different from
time to time for a given source, the geometric location
(as inferred from the kHz QPO frequency) of the disk
with respect to the neutron star does not seem to change by much.  This
is just another way of saying that the mass accretion rate does not
directly determine the QPO frequency.  It is the disk regime that
determines the large scale of the kHz QPO frequency and the mass
accretion rate affects it in the smaller scale in between regime
changes.

We concluding by noting that a possible scenario explaining the observed two
regimes on the QPO frequency vs. flux plane is that the disk structure
changed during the decay of the outburst. The different regimes
correspond  to different vertical scale heights. Therefore kHz QPOs
can be used to probe  the accretion disk structure evolution.

\acknowledgments

We thank the anonymous referee for constructive criticisms and
suggestions that have helped us improve this paper.  We thank P. Ghosh
and W. Yu for useful discussions.

\pagebreak

\pagebreak

\pagebreak

\pagebreak

\pagebreak

\end{document}